\documentclass{optica-article}

\journal{opticajournal} 

\articletype{Research Article}

\usepackage{cite}
\usepackage{multirow}
\usepackage{amsmath}
\usepackage{upgreek}
\usepackage{makecell}
\usepackage{lineno}

\begin{document}

\title{Toward ground-truth optical coherence tomography via three-dimensional unsupervised deep learning processing and data}

\author{Renxiong Wu,\authormark{1} Fei Zheng,\authormark{1} Meixuan Li,\authormark{1} Shaoyan Huang,\authormark{1} Xin Ge,\authormark{2} Linbo Liu,\authormark{3} Yong Liu,\authormark{1} and Guangming Ni\authormark{1,*}}

\address{\authormark{1}School of Optoelectronic Science and Engineering, University of Electronic Science and Technology of China, Chengdu 610054, China\\
\authormark{2}SchooL of Science, Shenzhen Campus of Sun Yat-sen University, Shenzhen 510275, China\\
\authormark{3}School of Electrical and Electronic Engineering, Nanyang Technological University, Singapore 639798, Singapore}

\email{\authormark{*}guangmingni@uestc.edu.cn} 


\begin{abstract*} 
Optical coherence tomography (OCT) can perform non-invasive high-resolution three-dimensional (3D) imaging and has been widely used in biomedical fields, while it is inevitably affected by coherence speckle noise which degrades OCT imaging performance and restricts its applications. Here we present a novel speckle-free OCT imaging strategy, named toward-ground-truth OCT (\emph{t}GT-OCT),  that utilizes unsupervised 3D deep-learning processing and leverages OCT 3D imaging features to achieve speckle-free OCT imaging. Specifically, our proposed \emph{t}GT-OCT utilizes an unsupervised 3D-convolution deep-learning network trained using random 3D volumetric data to distinguish and separate speckle from real structures in 3D imaging volumetric space; moreover, \emph{t}GT-OCT effectively further reduces speckle noise and reveals structures that would otherwise be obscured by speckle noise while preserving spatial resolution. Results derived from different samples demonstrated the high-quality speckle-free 3D imaging performance of \emph{t}GT-OCT and its advancement beyond the previous state-of-the-art. 

\end{abstract*}

\section{Introduction}
\label{sec:introduction}
Optical coherence tomography (OCT) stands as a pivotal noninvasive biomedical imaging technology and can produce three-dimensional images of various biological tissues with micrometer resolution \cite{Three‐dimensional,Speckle-modulating}. Beyond its established utility in ophthalmology, OCT has progressively gained popularity in clinical diagnostic applications across cardiology, dermatology, and other fields \cite{SingleA-Line, GuangmingBook}. However, an intrinsic challenge of OCT lies in the presence of speckle noise within OCT images, which noticeably compromises their quality and consequently impairs subsequent interpretation and diagnosis.

Many speckle-reduction methods for OCT imaging have been proposed; among these, speckle-modulating OCT (SM-OCT) is a well-regarded method \cite{Speckle-modulating}. By using a moving diffuser, SM-OCT can acquire an unlimited number of uncorrelated speckle patterns and effectively remove speckle noise without degrading the spatial resolution of the images, which helps SM-OCT clarify and reveal structures that are otherwise obscured or undetectable. However, SM-OCT uses a moving diffuser in the optical path and has to perform repeated B-scans, which substantially reduces the imaging sensitivity and temporal resolution of the OCT system.

Recently, methods that leverage the robust data-fitting capabilities of deep learning, have been widely used for OCT despeckling. With the rapid development of deep learning-based methods, two categories have emerged: those that do and do not require clean images\cite{Triplet}. The former category encompasses both supervised and unpaired methods. Supervised denoising methods necessitate paired noisy and clean images for network training. DnCNN \cite{GaussianDenoiser}, a method based on residual learning has been proposed to achieve denoising performance. Additionally, the generative adversarial network (GAN) has gained immense popularity for supervised OCT despeckling methods, giving rise to approaches such as Caps-cGAN \cite{Semi-supervised}, SDSR-OCT \cite{Simultaneous}, DNGAN \cite{DN-GAN}, SiameseGAN \cite{SiameseGAN}, and MDR-GAN \cite{multi-scale} that have outperformed conventional CNNs in enhancing image quality. Sm-Net OCT \cite{Sm-NetOCT} involves training a GAN with a customized SM-OCT dataset to despeckle and resolve intricate structures. For unpaired methods, unpaired clean and noisy images are needed to train a cycleGAN-based network, such as HDcycleGAN \cite{DomainShift}, SPcycleGAN \cite{WuNoise}, DRGAN \cite{disentangled}, and ADGAN \cite{ADGAN}. These unpaired methods disentangle OCT images into content and noise domains, approximating the denoising performance of supervised methods. Nonetheless, the dependency of these methods on clean images, which are often challenging to acquire due to repeated lengthy scanning procedures, motivates exploration into methods that do not require clean images.

Consequently, several methods that do not require clean images have been explored to address this concern. The Noise2Noise (N2N) strategy \cite{Consensus}, using pairs of noisy images of the same scene, has been applied to OCT despeckling \cite{Real-time,Comparative,N2NSR‐OCT}. Nevertheless, the reliance of this approach on at least two scans of the same sample location remains a practical barrier. Furthermore, there are single-image denoising techniques that are notable in natural image processing; these includes Noise2Void (N2V) \cite{Noise2void}, Noise2Self (N2S) \cite{Noise2self}, Neighbor2Neighbor (NBR) \cite{Neighbor2Neighbor}, etc. In OCT despeckling, the use of NBR, a method  with a multiscale pixel patch sampler \cite{Self-supervised} ensures both despeckling efficacy and structure preservation. The application of Transformer \cite{Unsupervisedinter-patch,transformer} has also been attempted to reduce speckle noise sufficiently and preserve details. These proposed methods represent great efforts to alleviate the problem of detail structure damage in single-image denoising methods but show limited improvement. To resolve finer biological detail structures, Noise2Context \cite{Noise2Context}, Noise2Stack \cite{Noise2Stack}, and Noise2Sim \cite{Similarity-based} have been proposed to augment denoised image details by exploiting shared information from adjacent noisy images within 3D volumetric data. However, reliance solely on short-distance adjacent slices fails to efficiently utilize all of the available 3D OCT data and has limited despeckling performance, especially in non-ophthalmological OCT imaging applications.

Here we present a novel strategy, called to toward-ground-truth OCT (\emph{t}GT-OCT), to achieve speckle-free OCT imaging, that is based on OCT 3D imaging features and unsupervised 3D deep-learning processing. By distinguishing and extracting speckle and structures in random OCT 3D imaging volumetric data using 3D-convolutional neural networks (3D-CNNs), the proposed \emph{t}GT-OCT effectively reduces speckle noise and reveals structures that are otherwise obscured or undetectable while preserving spatial resolution. Qualitative and quantitative results for various 3D OCT images including those of human retina, other human tissues, meats, and Scotch tape demonstrate the state-of-the-art performance of \emph{t}GT-OCT in despeckling, even achieving microstructure resolution performance comparable to that of SM-OCT, which is regarded as the gold standard for OCT despeckling. Meanwhile, our work has also provided a new perspective for studying OCT speckle-free imaging by utilizing the 3D imaging characteristics of OCT alongside unsupervised 3D deep learning processing.

\section{METHODS}

\subsection{3D unsupervised deep learning extracts OCT speckle patterns in 3D space}

\begin{figure}[ht]
\centering
\includegraphics[width=0.8\linewidth]{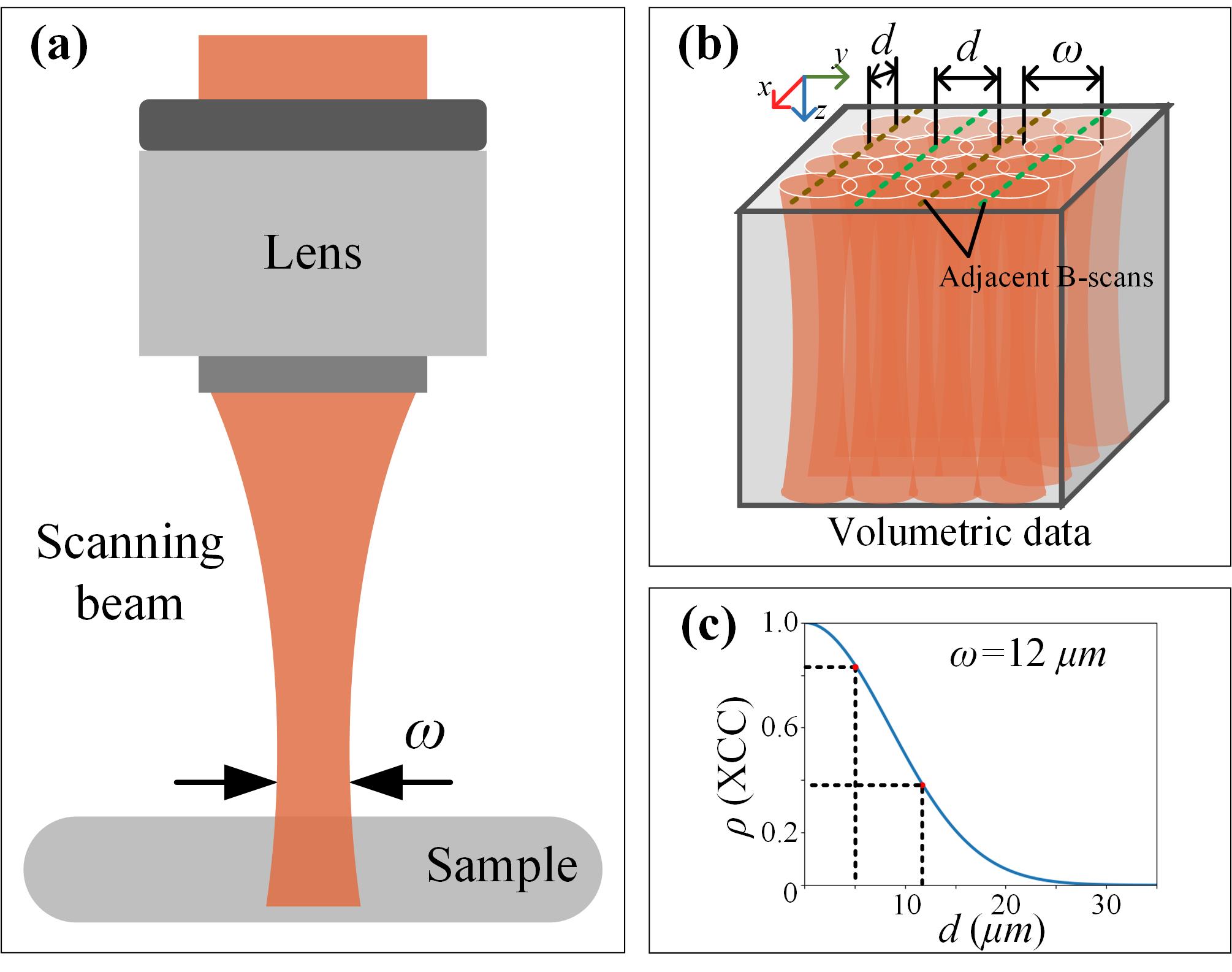}
\caption{The correlation of adjacent A-scans depends on the limited beam size and appropriate lateral displacement. (a) The scanning beam size; (b) volumetric data containing uncorrelated and weakly correlated speckle patterns and strongly correlated sample structures when the displacement d is smaller than beam size; (c) the relationship between XCC and lateral displacement.}
\label{fig:principle}
\end{figure}

Speckle-modulating OCT uses a moving diffuser and requires repeated scanning at the sample to acquire a large number of uncorrelated speckle patterns and then performs an averaging operation to effectively remove speckle noise without degrading the spatial resolution of the images. Our previously proposed method, Sm-Net OCT\cite{Sm-NetOCT},  uses deep-learning network to distinguish and extract those large number of uncorrelated speckle patterns in SM-OCT speckle images with SM-OCT speckle-free images (used as ground truth) to generate speckle-free OCT images. 
OCT speckle patterns depend on scanning voxel sizes and sample structures \cite{Speckle-modulating,Speckle}, so OCT 3D volumetric data can contain unlimited uncorrelated speckle patterns; moreover,  these speckle patterns can potentially be distinguished and separated in 3D space for OCT speckle-free imaging. Meanwhile, neighboring B-scans also contain mass strongly correlated sample structures in OCT 3D volumetric data, as shown in Fig. \ref{fig:principle}.

In OCT 3D imaging, the correlation of two A-scans can be expressed using the Pearson cross-correlation coefficient (XCC) \cite{Robustmotion}. Importantly, the XCC between two adjacent slightly displaced A-scans has an explicit functional dependency on the  lateral distance d and can be expressed as (\ref{eq:refname1}), where $\omega $, is the Gaussian beam waist of the light beam, which is also the transverse optical resolution of the OCT system.
\begin{equation}
\rho  = \exp \left( { - \frac{{{d^2}}}{{{\omega ^2}}}} \right)
\label{eq:refname1}
\end{equation}
As shown in Figs. \ref{fig:principle}(b) and (c), owing to the limited scanning volume size of OCT, adjacent A-scans can contain both uncorrelated and weakly correlated speckle patterns at suitably minor lateral displacements $d$, while neighboring B-scans can have strongly correlated structures \cite{Depth-resolved}. Therefore, OCT 3D volumetric data can contain mass uncorrelated speckle patterns and strongly-correlated structural information. The strongly-correlated structural information can further act as the ground truth for our proposed unsupervised 3D deep learning network described in the following sections; this ground truth helps our 3D deep learning network to further distinguish and extract speckle patterns and generate speckle-free OCT images.

\begin{figure}[ht]
\centering
\includegraphics[width=\linewidth]{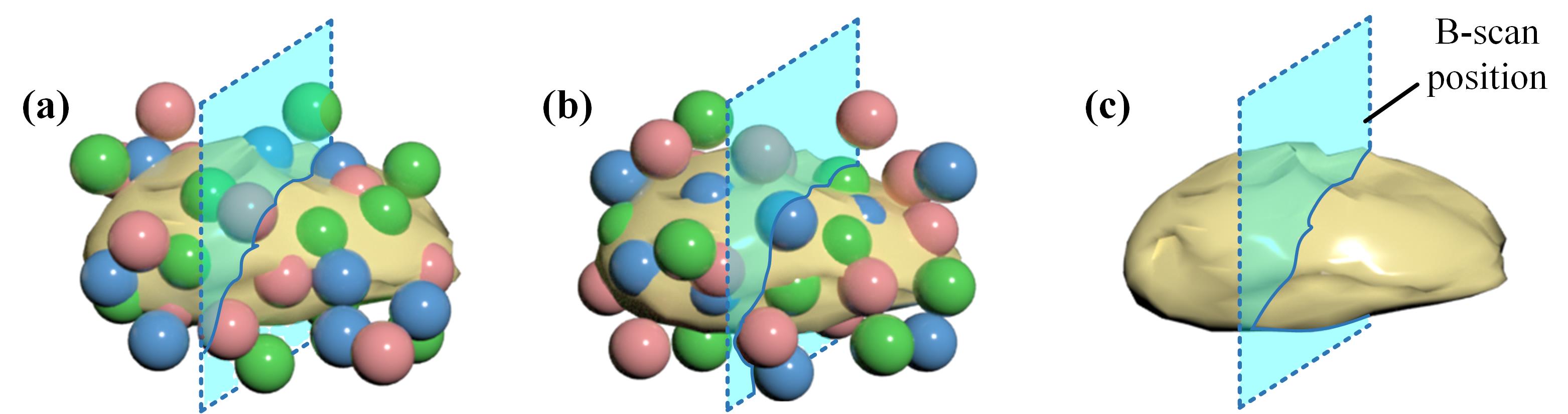}
\caption{Schematic of clean volumetric data and paired noisy volumetric data. (a) and (b) are the paired noisy volumes with uncorrelated speckle patterns, (c) is the clean volume.}
\label{fig:3Dmax}
\end{figure}

Unsupervised networks \cite{Consensus,Neighbor2Neighbor} analyzing two similar (strongly correlated sample structures) but different noisy (uncorrelated speckle patterns) 2D images have been proposed to reduce noise. Here we extend this principle to the realm of 3D volumetric data. First we consider two noisy volumetric image sets $y=[y_0,y_1,\cdots,\ y_{n-1}]$ and $z=[z_0,z_1,\cdots,\ z_{n-1}]$, which are sequences of n multiple B-scans. The noisy data y and z are independent conditional on the clean data $x=[x_0,x_1,\cdots,x_{n-1}]$. We aim to train a 3D network parametrized by by minimizing the function as (\ref{eq:refname2}). Fig.  \ref{fig:3Dmax} shows the schematic of the clean data $x$ and noisy data $y$ and $z$ with substantial uncorrelated speckle patterns.
\begin{equation}
\mathop {\arg {\rm{ }}\min }\limits_\theta  \mathbb{E}{\left\| {{f_\theta }\left( y \right) - z} \right\|^2}
\label{eq:refname2}
\end{equation}
Assume that ${f_\theta }\left( y \right) = x$ and ${f_\theta }\left( z \right) = x + \varepsilon $, where $\varepsilon  \ne 0$, (\ref{eq:refname3}) can be expressed,
\begin{equation}
\mathbb{E}{\left\| {{f_\theta }\left( y \right) - z} \right\|^2} = \mathbb{E}{\left\| {{f_\theta }\left( y \right) - x} \right\|^2} + \sigma _z^2 - 2\varepsilon \mathbb{E}\left( {{f_\theta }\left( y \right) - x} \right)
\label{eq:refname3}
\end{equation}
where $\mathbb{E}{\left\| {{f_\theta }\left( y \right) - x} \right\|^2}$ represents the loss function of the supervised learning algorithm using clean images, and $\sigma_z^2$ is a constant. Note that minimizing $\mathbb{E}{\left\| {{f_\theta }\left( y \right) - z} \right\|^2}$ converges to minimize $\mathbb{E}{\left\| {{f_\theta }\left( y \right) - x} \right\|^2}$ when gap $\varepsilon  \to 0$. This means that the label for the training network can be noisy data that are similar to the input data rather than a clean data.

\subsection{Generation of paired 3D volumetric training data}
\begin{figure}[ht]
\centering
\includegraphics[width=\linewidth]{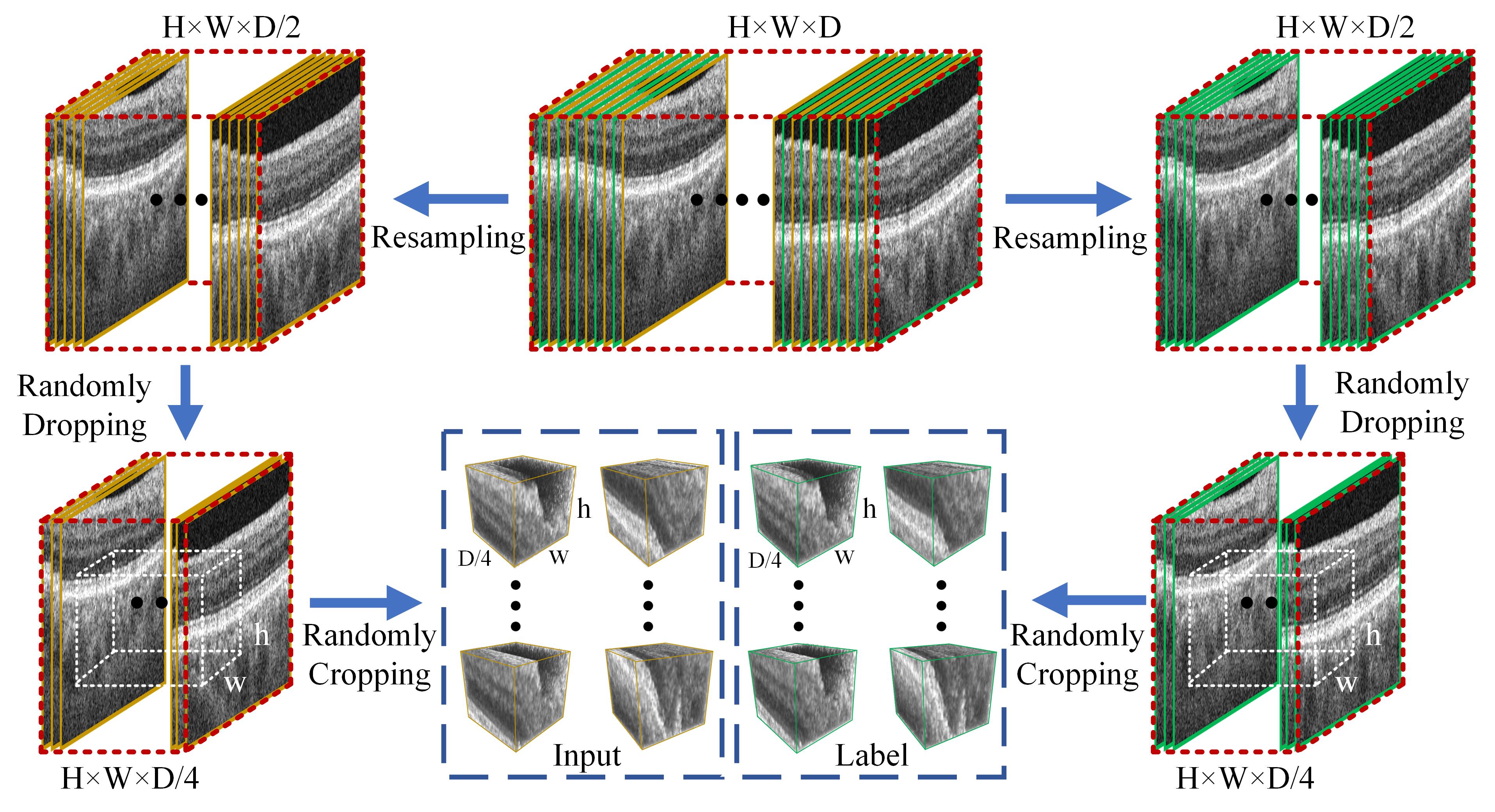}
\caption{The pipeline of generating paired 3D volumetric data. The original volumetric data are resampled adjacently to two similar volumetric data and then divided into paired volumetric blocks by randomly dropping and cropping.}
\label{fig:3Ddata}
\end{figure}

Here, we prepared paired 3D volumetric data that were similar but not identical for training our unsupervised network; pairs were generated by processing one OCT 3D volumetric dataset with strongly correlated sample structures and uncorrelated speckle patterns at suitably minor lateral displacements.Fig.  \ref{fig:3Ddata} shows the processing pipeline for generating paired data. Specifically, the noisy 3D data with width $W$, height $H$ and depth $D$ are processed using three steps: (1) The 3D data is divided into two sub-datasets by selecting adjacent B-scans along the lateral direction. One subdata point consists of the (2i-1)-th B-scan while the other subdata point consists of the 2i-th B-scan from the original 3D data, where $i=1,2\cdots,D/2$. (2) B-scans in the subdata are randomly dropped, and in experiment, either one of two B-scans or two of four B-scans are dropped without repetition. (3) The paired 3D input and label are generated by random cropping the images with a cropping block of size (h, w, D/4). As neighboring B-scans in the original 3D data contain strongly correlated structures, resampling and dropping operations ensure that the paired volumetric data retain similar sample structures but different speckle patterns.

\subsection{Three-dimensional datasets}
Here we employed two customized volumetric datasets for both training and testing, as well as two additional volumetric datasets used exclusively for testing. Detailed characteristics of these datasets are summarized in Table \ref{tab:dataset}. The data for the OCT-R1 dataset were collected from 41 human eyes using a BM-400K BMizar (Topi Ltd.) scanner at Sichuan Provincial People’s Hospital (IRB-2022-258). To enhance the diversity of the data, we conducted scans over two different ranges. For the OCT-N1 dataset, we used a customized OCT setup \cite{Three‐dimensional} to collect 46 three-dimensional data points from different samples, including Scotch tape, pork, human skin, and a placenta. The test datasets were only used for the evaluation phase, and the network was never exposed to these data during training. The OCT-R2 dataset includes five three-dimensional images of the human retina acquired using Spectralis OCT (Heidelberg Engineering Inc.). The OCT-N2 dataset contains 3D noisy and clean data of samples such as Scotch tape, fish and pork, which were collected by our SM-OCT setup \cite{Speckle-modulating}. For generation of clean data references using SM-OCT, we shifted the optical diffuser and simultaneously scanned the same position for 50 times. The datasets are available for download at \underline{https://tianchi.aliyun.com/dataset/161472}.
\begin{table*}[ht]
\centering
\caption{The details of training and test datasets}
\resizebox{\textwidth}{!}{
\begin{tabular}{lllllll}
\hline
\makebox[0.05\textwidth][l]{Dataset}                & \makebox[0.05\textwidth][l]{OCT Setup}                       & \makecell[l]{Axial \\ resolution} & \makecell[l]{Lateral \\ displacement} & \makebox[0.05\textwidth][l]{Data size}      & \makebox[0.05\textwidth][l]{Subject} & \makebox[0.05\textwidth][l]{Application}                                                                                \\ \hline
\multirow{2}{*}{OCT-R1} & \multirow{2}{*}{BM-400K BMizar} & \multirow{2}{*}{$\sim$3.8 mm}                               & \multirow{2}{*}{10 mm}                                          & 512×512×512      & 25      & \multirow{2}{*}{\begin{tabular}[c]{@{}l@{}}T (Training) \& \\ V (Validating)\end{tabular}} \\ \cline{5-6}
                        &                                 &                                                             &                                                                 & 1948×1536×1280 & 16      &                                                                                            \\
OCT-N1                  & Customized OCT                  & $\sim$1.68 mm                                               & 5.10 mm                                                         & 800×800×800    & 46      & T \& V                                                                                     \\
OCT-R2                  & Spectralis OCT                  & $\sim$4.1 mm                                                & 11.8 mm                                                         & 512×512×400    & 5       & Testing                                                                                    \\
OCT-N2                  & Customized SM-OCT               & $\sim$1.68 mm                                               & 5.10 mm                                                         & 800×420×100    & 4       & Testing                                                                                    \\ \hline
\end{tabular}
}
\label{tab:dataset}
\end{table*}

\subsection{Network}
\begin{figure*}[ht]
\centering
\includegraphics[width=\linewidth]{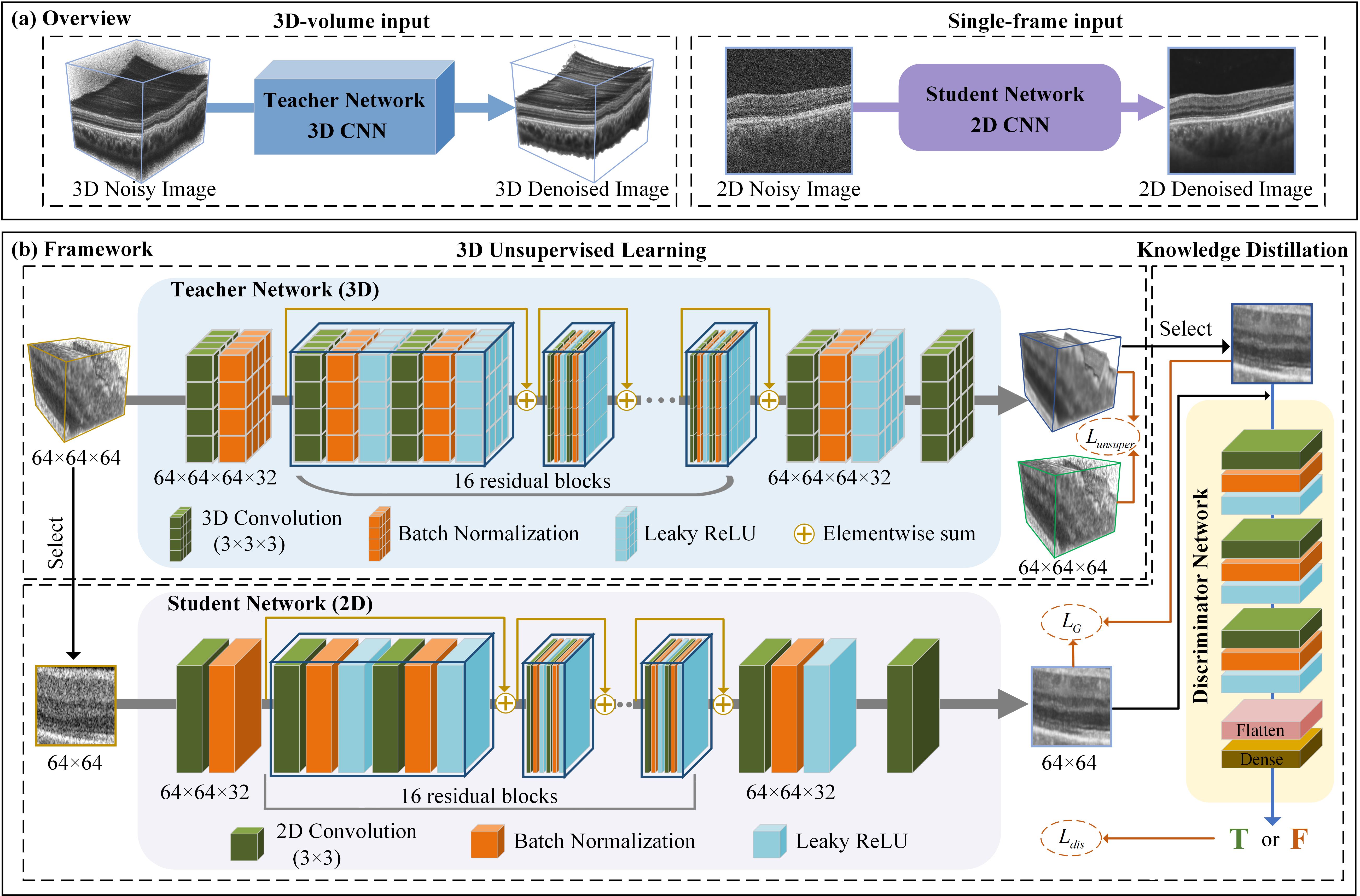}
\caption{\centering{Proposed 3D network and distillation architecture. (a) Overview, and (b) training framework.}}
\label{fig:network}
\end{figure*}

Fig.  \ref{fig:network} illustrates the architecture of our novel \emph{t}GT-OCT strategy. This comprehensive framework consists of both a 3D CNN and a 2D CNN to accommodate OCT data with varying dimensions, i.e., for 3D volumetric inputs and single-frame inputs respectively. The 3D CNN is trained with an unsupervised learning strategy that uses paired similar noisy volumetric data: one of which is considered the input while the other is the target; the mean square error is used as the loss function. We introduce a knowledge distillation mechanism to distill the knowledge of the 3D CNN (as a teacher network) into the 2D CNN (as a student network) to adapt to single image denoising and save computing resources. As the teacher network, the well-trained 3D CNN yields denoised volumetric data, from which the corresponding denoised frame is chosen as the label for training the student network. The discriminator network is used for comparison against the student network to allow it to generate a denoised output that is closer to the output of the 3D CNN. Depending on whether a 2D single-frame input or a 3D volumetric input is encountered, the teacher and student networks are employed as necessary components of the inference stage.

\subsubsection{Unsupervised 3D CNN}
The presented 3D CNN mentioned above is an evolution of the deep residual network (ResNet), which is commonly used for image super-resolution and denoising. As shown in Fig. \ref{fig:network}(b), ResNet consists of pre-residual layer, sixteen residual blocks and a post-residual layer. The pre-residual layer was a 3D convolutional layer and is followed by a batch normalization layer. The 3D convolutional layer has 32 filters, each with $\mathrm{3\times3\times3}$ kernel size. Each of the residual blocks has two 3D convolutional layers that are identical to pre-residual layer, followed by batch normalization layers; the network uses with the leaky ReLU function as the activation layer. The post-residual layer consists of a $\mathrm{3\times3\times3}$ 3D convolutional layer, batch normalization layer, leaky ReLU function and a $\mathrm{1\times1\times1}$ 3D convolutional layer. A skip connection is introduced in each residual block to connect the input and output of the block. The size of the feature images passed in the network remains invariant, sustaining a size of $\mathrm{64\times64\times64}$ during the training phase. 

\subsubsection{Knowledge distillation}
In recognition of the resource-intensive nature of three-dimensional convolution and the existence of OCT dataset that only include single noisy images, we introduced a knowledge distillation mechanism to distill the knowledge of the 3D CNN into a 2D CNN. As shown in Fig. \ref{fig:network}(b), we denoted the well-trained 3D CNN as the teacher network and the 2D CNN as the student network. One noisy frame and corresponding denoised frame were selected from noisy 3D volumetric data and the denoised 3D volumetric output of the teacher network, respectively. These frames were used to train the student network with supervised learning strategy. To retain as many feature maps as possible during knowledge distillation, the structure of the student network was aligned with the teacher network, changing 3D convolution layers in the teacher network to 2D convolution layers. All convolutional layers are $\mathrm{3\times3}$ in size except for the last layer whose size is $\mathrm{1\times1}$.

\subsubsection{Adversarial learning}
The generative adversarial network (GAN) consists of a generator and a discriminator; such networks are widely used in image-to-image translation. The generator network is optimized to generate a high-quality denoised image close to the label that would be indistinguishable by the discriminator. Meanwhile, the discriminator network is also enhanced to distinguish between labels and images generated by the generator. Therefore, adversarial learning allows the generator network to achieve better performance compared to using only generative models. Here, adversarial training between the student network and the discriminator network was conducted. Specifically, the discriminator network was tasked with discerning the authenticity of the output image from the student network when juxtaposed against the ground truth reference. For the discriminator network, convolutional layers with a $\mathrm{3\times3}$ kernel size followed by a batch normalization layer were utilized for all layers. The strides of the convolutional layers were $\mathrm{2\times2}$ and a leaky ReLU function was used as the activation layer. A flattened layer, a dense layer, and a sigmoid function were used to determine the classification probability.

\subsubsection{Objective functions}
The unsupervised learning process of the 3D CNN follows the principle of the Noise2Noise method, the objective function is defined as (\ref{eq:refname4}):
\begin{equation}
{L_{unsuper}} = \frac{1}{{HWF}}\sum\limits_{i,j,k = 1}^{H,W,F} {{{\left\| {{v_{i,j,k}} - {p_{i,j,k}}} \right\|}^2}}
\label{eq:refname4}
\end{equation}
where $v_{i,j,k}$ and $p_{i,j,k}$ are pixels of two generated similar noisy OCT volumetric data and $\mathrm{H,\ W,\ and\ F}$ are the height, width, and frames of the volumetric data, respectively.
When training the student network, we optimize a GAN with objective functions that can be described as (\ref{eq:refname5}) and (\ref{eq:refname6}):
\begin{equation}
\mathop {\min }\limits_G Los{s_{adv}}\left( G \right) = \frac{1}{{HW}}\sum\limits_{i,j = 1}^{H,W} {\log \left( {1 - D\left( {G\left( {{x_{i,j}}} \right)} \right)} \right)}
\label{eq:refname5}
\end{equation}

\begin{equation}
\begin{aligned}
\mathop {\min }\limits_D Los{s_{dis}}\left( D \right) &= \frac{1}{{HW}}\sum\limits_{i,j = 1}^{H,W} {\log \left( {D\left( {{y_{i,j}}} \right)} \right)} \\
& + \frac{1}{{HW}}\sum\limits_{i,j = 1}^{H,W} {\log \left( {D\left( {G\left( {{x_{i,j}}} \right)} \right) - 1} \right)}     
\end{aligned}
\label{eq:refname6}
\end{equation}
where $G$ and $D$ are generator and discriminator networks, respectively, and $x_{i,j}$ and $y_{i,j}$ are pixels of the noisy image and ground truth, respectively.
Using mean square error (MSE) loss in generator loss is beneficial to improve image quality and previous work showed that using VGG loss can maintain high-frequency information and structure details. VGG loss is the Euclidean distance between the high-level perceptual features of the generated image and ground truth extracted by the VGG network pre-trained on the ImageNet dataset. The MSE loss and VGG loss are expressed as  (\ref{eq:refname7}) and (\ref{eq:refname8}):
\begin{equation}
{L_{MSE}} = \frac{1}{{HW}}\sum\limits_{i,j = 1}^{H,W} {{{\left\| {{x_{i,j}} - {y_{i,j}}} \right\|}^2}} 
\label{eq:refname7}
\end{equation}

\begin{equation}
{L_{Vgg}} = \frac{1}{{HW}}\sum\limits_{i,j = 1}^{H,W} {{{\left\| {VG{G_{16}}{{\left( x \right)}_{i,j}} - VG{G_{16}}{{\left( y \right)}_{i,j}}} \right\|}^2}} 
\label{eq:refname8}
\end{equation}
where ${VGG}_{16}$ represents the VGG-16 network. Hence, the generator network minimized the combination of MSE, Vgg, and adversarial loss, and is described as (\ref{eq:refname9}):
\begin{equation}
{L_G} = \alpha {L_{adv}} + \beta {L_{MSE}} + \gamma {L_{Vgg}}
\label{eq:refname9}
\end{equation}
where $\alpha$, $\beta$, $\gamma$ are weight coefficients of loss term.

\begin{figure*}[ht]
\centering
\includegraphics[width=0.9\linewidth]{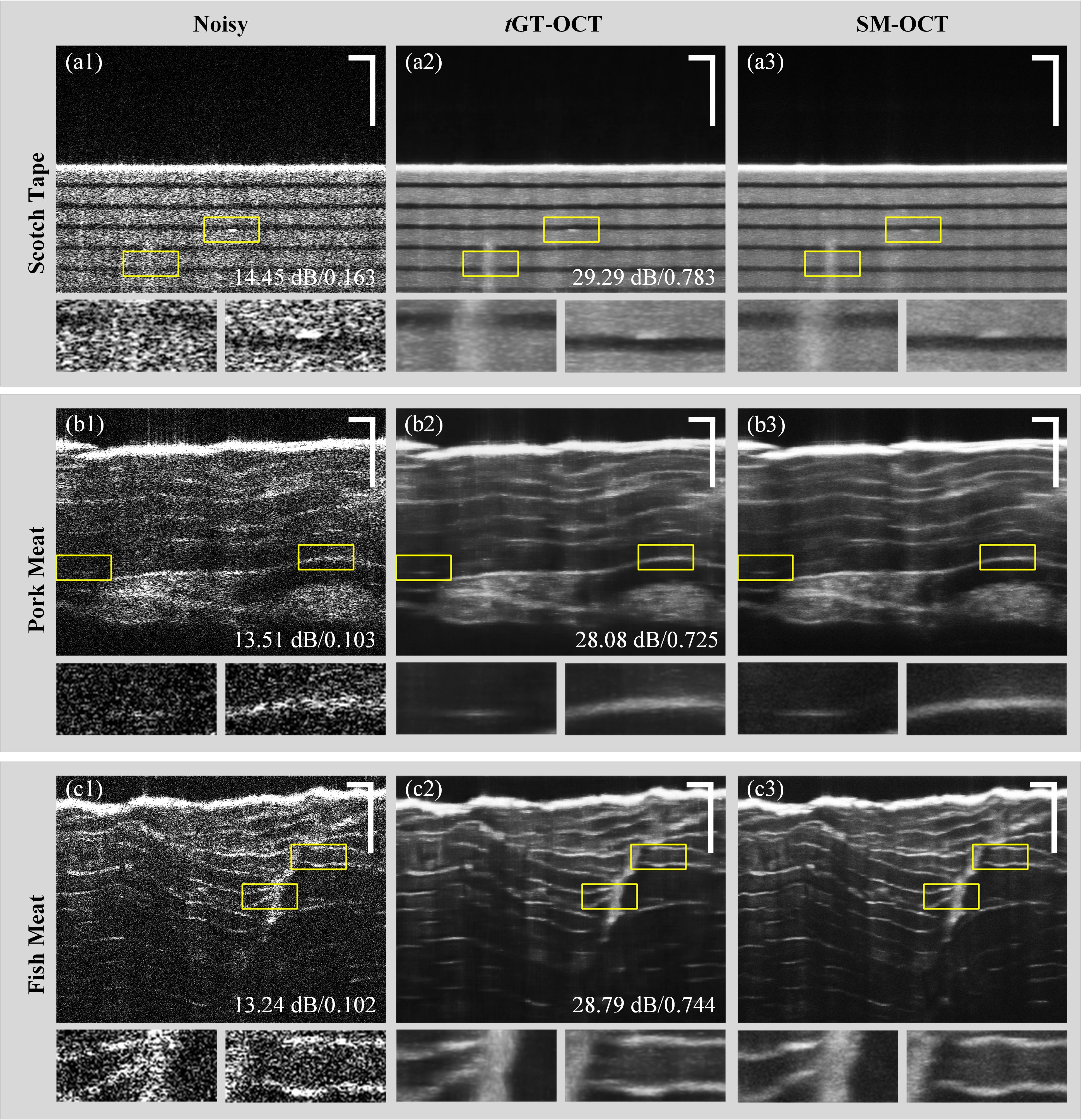}
\caption{Image outputs of the proposed \emph{t}GT-OCT method, and comparison with SM-OCT images of Scotch tape, pork and fish. The evaluation metrics format is PSNR/SSIM. Scale bar: 200 $\upmu$m.}
\label{fig:ReSM}
\end{figure*}
\section{experiment and results}
\subsection{Experimental setup}
We implemented the proposed \emph{t}GT-OCT 3D deep learning network in the TensorFlow framework. Training was conducted on a platform equipped with an Intel Xeon Silver 4210R CPU, an NVIDIA Quadro RTX6000 graphics card boasting 48 GB of memory, and a total of 64 GB RAM. Adam was adopted as the optimizer with the momentum 1 and momentum 2 parameters set to 0.5 and 0.999, respectively. The learning rate was set to $5 \times {10^{ - 5}}$. When training the teacher network, the batch size and number of iterations were set to 4 and 100,000, respectively. The student network underwent training with a batch size of 8 and a cumulative iteration count of 150,000. The weights $\alpha$, $\beta$, $\gamma$ in  (\ref{eq:refname9}) were empirically set to 0.001, 1 and 1 according to the results of several experiments. We trained two models using the OCT-R1 and OCT-N1 datasets. To expedite training and streamline computational resource utilization, the length, width and height of the volumetric data were randomly cropped to $\mathrm{64\times64\times64}$ and the 2D single-frame OCT images were cropped to $\mathrm{480\times480}$.

When there was no ground truth, signal-to-noise ratio (SNR)\cite{evaluation}, contrast-to-noise ratio (CNR)\cite{evaluation} and equivalent number of looks (ENL)\cite{evaluation} were utilized to evaluate the denoising performance. SNR indicates the radio of signal energy and noise energy, CNR measures the contrast between a feature of interest and background noise and ENL measures smoothness in homogeneous areas.
When comparing model outputs with the ground truth, the structure similarity (SSIM)\cite{evaluation} index and peak signal-to-noise ratio (PSNR)\cite{evaluation} were also used. They are calculated respectively to compare structural similarity and signal energy with ground truth.

\subsection{Experimental results}
\subsubsection{Comparison with SM-OCT}

\begin{table}[ht]
\centering
\caption{Quantitative results of the proposed \emph{t}GT-OCT compared with SM-OCT}
\begin{tabular}{lllllll}
\hline
\multicolumn{1}{l}{}                              &                              & PSNR   & SSIM  & SNR    & CNR    & ENL     \\ \hline
\multicolumn{1}{c|}{\multirow{2}{*}{\makecell[l]{Scotch \\Tape}}} & Noisy   & 14.451 & 0.163 & 15.546 & 1.684  & 0.360   \\
\multicolumn{1}{c|}{}                             & \emph{t}GT-OCT & 29.286 & 0.783 & 39.420 & 6.142  & 118.076 \\ \hline
\multicolumn{1}{c|}{\multirow{2}{*}{\makecell[l]{Pork \\Meat}}}   & Noisy   & 13.510 & 0.103 & 10.847 & -0.305 & 0.320   \\
\multicolumn{1}{c|}{}                             & \emph{t}GT-OCT & 28.078 & 0.725 & 40.151 & 3.094  & 574.779 \\ \hline
\multicolumn{1}{c|}{\multirow{2}{*}{\makecell[l]{Fish \\Meat}}}   & Noisy   & 13.239 & 0.102 & 5.114  & -3.098 & 0.402   \\
\multicolumn{1}{c|}{}                             & \emph{t}GT-OCT & 28.788 & 0.744 & 30.917 & 0.603  & 139.853 \\ \hline
\end{tabular}
\label{tab:metricSM}
\end{table}

Speckle-modulating OCT can be used to acquire the gold-standard ground truth image for OCT despeckling, and is widely used in supervised deep learning methods \cite{Sm-NetOCT,Hybrid-structure,Dong}. 
To demonstrate the performance of our proposed \emph{t}GT-OCT, we trained the \emph{t}GT-OCT 3D deep learning network on the OCT-N1 dataset and tested it on the OCT-N2 dataset by comparing the outputs with SM-OCT images. Fig.  \ref{fig:ReSM} provides a comprehensive visual representation, featuring the original noisy images, \emph{t}GT-OCT images, and the corresponding SM-OCT images of Scotch tape, pork, and fish samples. As shown in Fig. \ref{fig:ReSM}(a1) - (a3), the magnified area on the right reveals a subtle structure which is comparable in size to speckle noise, concealed within the noisy image. In contrast, this structural information is clearly visible in the \emph{t}GT-OCT image, which is consistent with the SM-OCT performance. In fish and pork images, the stripes are rendered distinctly in the \emph{t}GT-OCT images, aligning closely with the SM-OCT images. Quantitatively, Table \ref{tab:metricSM} presents the quantitative results of noisy and denoised images by \emph{t}GT-OCT.  The PSNR scores for \emph{t}GT-OCT images of Scotch tape, pork and fish images are 29.286 dB, 28.078 dB and 28.788 dB, respectively, exhibiting substantial improvement over their noisy counterparts. The SSIM scores of \emph{t}GT-OCT images are also significantly improved, reaching 0.783, 0.725 and 0.744 respectively. SNR, CNR and ENL of \emph{t}GT-OCT results are improved obviously compared with noisy images. The results demonstrated that the proposed \emph{t}GT-OCT adeptly suppresses speckle noise while simultaneously preserving microstructures, similar to SM-OCT's capabilities, all without necessitating hardware modifications.

\begin{figure}[!h]
\centering
\includegraphics[width=0.9\linewidth]{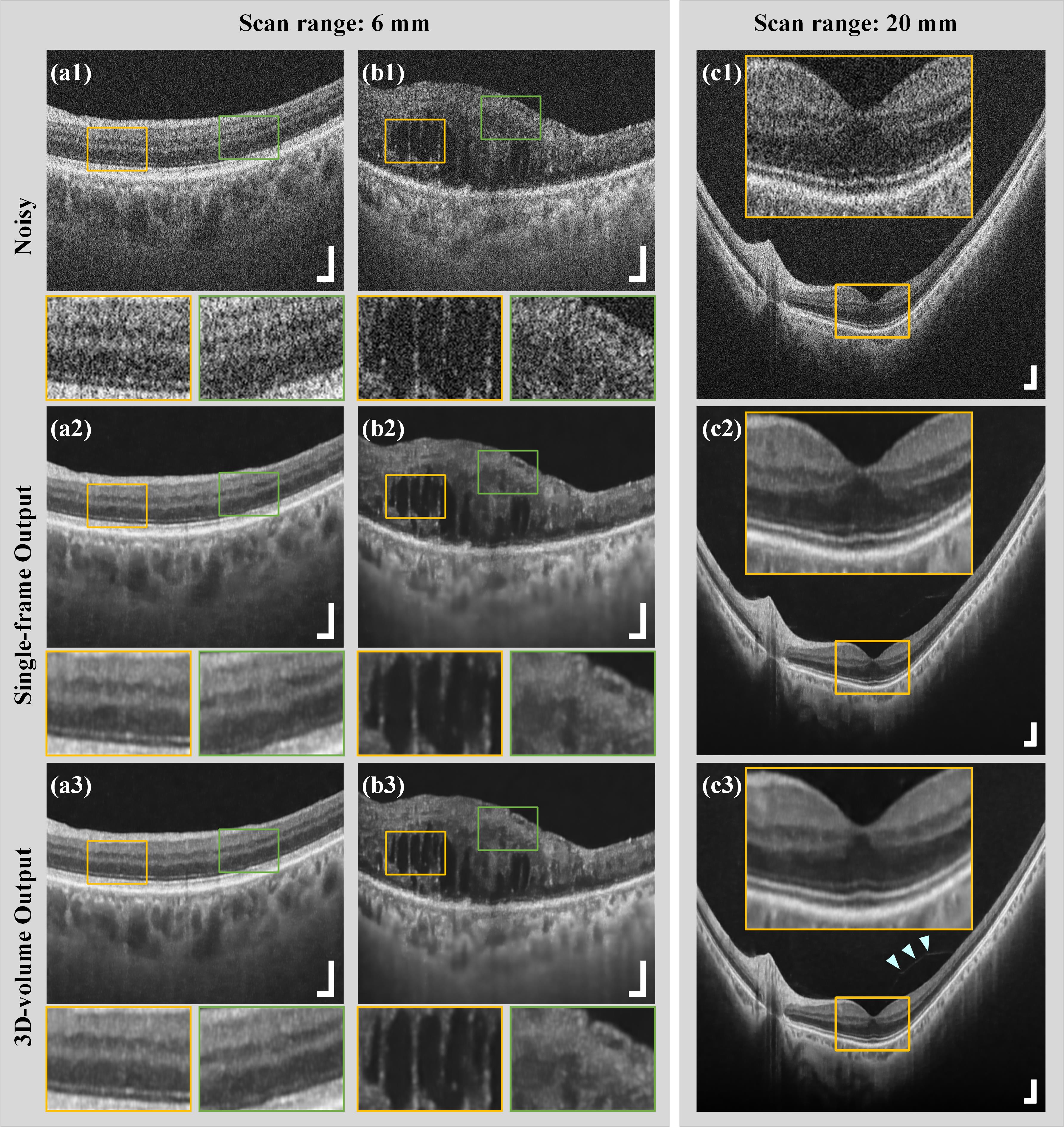}
\caption{Denoised images of the proposed \emph{t}GT-OCT on human retina data with different scan ranges. Scale bar: 200 $\upmu$m.}
\label{fig:ReRetina}
\end{figure}

\subsubsection{Clinical \emph{t}GT-OCT performance}
Here, we conducted a denoising experiment on clinical OCT images, including those of a human retina, human skin and human placenta. The \emph{t}GT-OCT teacher network (3D network) processed the 3D volumetric data, while the \emph{t}GT-OCT student network (2D network) handled the single frames to demonstrate the effectiveness of \emph{t}GT-OCT and knowledge distillation. As shown in Fig. \ref{fig:ReRetina}, the denoising process was applied to human retinal data with varied lateral scanning ranges (6 mm and 20 mm). Figs. \ref{fig:ReRetina}(a1) - (c1) show the noisy images, while Figs. \ref{fig:ReRetina}(a2) - (c2) and Figs. \ref{fig:ReRetina}(a3) - (c3) depict the \emph{t}GT-OCT results obtained via the \emph{t}GT-OCT 2D and 3D networks. 
These visual comparisons show the excellent speckle noise suppression and detail preservation capabilities of both networks, as well as their effectiveness across different scanning ranges and pathologies. The magnified regions shown also accentuate the enhanced distinctions between different layers of the retina, affirming the effectiveness of \emph{t}GT-OCT denoising. The \emph{t}GT-OCT 3D network outperforms its \emph{t}GT-OCT 2D counterpart by further resolving microstructures; this is possible because the 3D network can make full use of 3D spatial information in OCT volumetric data. In Fig. \ref{fig:ReRetina}(b1), the boundaries of the retinal edema cystic cavities in the magnified orange region are very difficult to distinguish as they are obscured by speckle noise. In Fig. \ref{fig:ReRetina}(b2), multiple cystic cavities can be distinguished after \emph{t}GT-OCT denoising, while the image shown in in Fig. \ref{fig:ReRetina}(b3), the septum of the cysts is not only clear but also have good continuity. Incomplete posterior vitreous detachment is seen at the point indicated by the red arrow in Fig. \ref{fig:ReRetina}(c3), and the septum produced by detachment can be clearly observed after being denoised by the \emph{t}GT-OCT 3D network, providing a clinical diagnosis.

\begin{figure}[!h]
\centering
\includegraphics[width=0.9\linewidth]{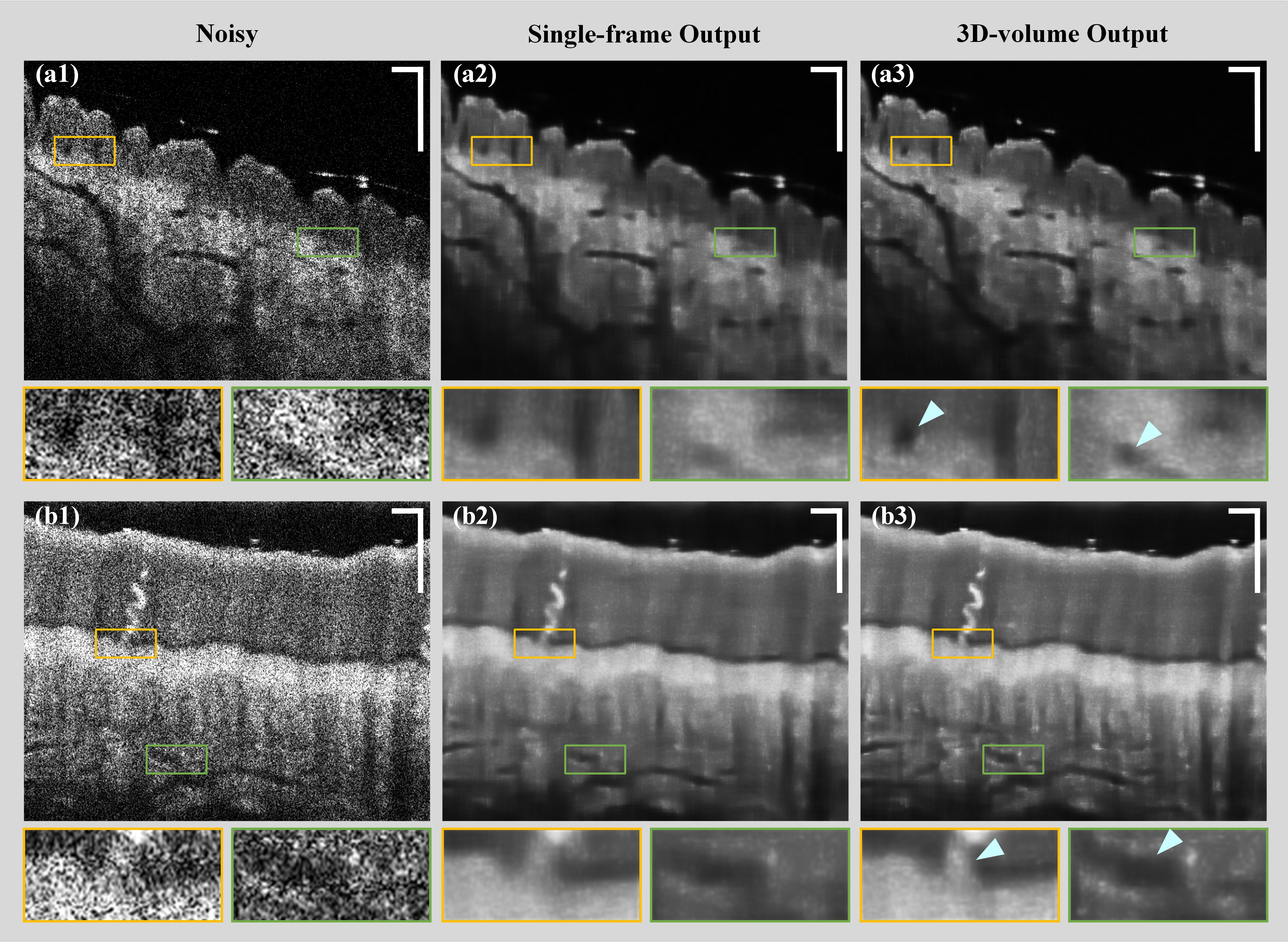}
\caption{Denoised images of the human skin output by the proposed \emph{t}GT-OCT. Scale bar: 200 $\upmu$m.}
\label{fig:ReSkin}
\end{figure}

\begin{figure}[!h]
\centering
\includegraphics[width=0.9\linewidth]{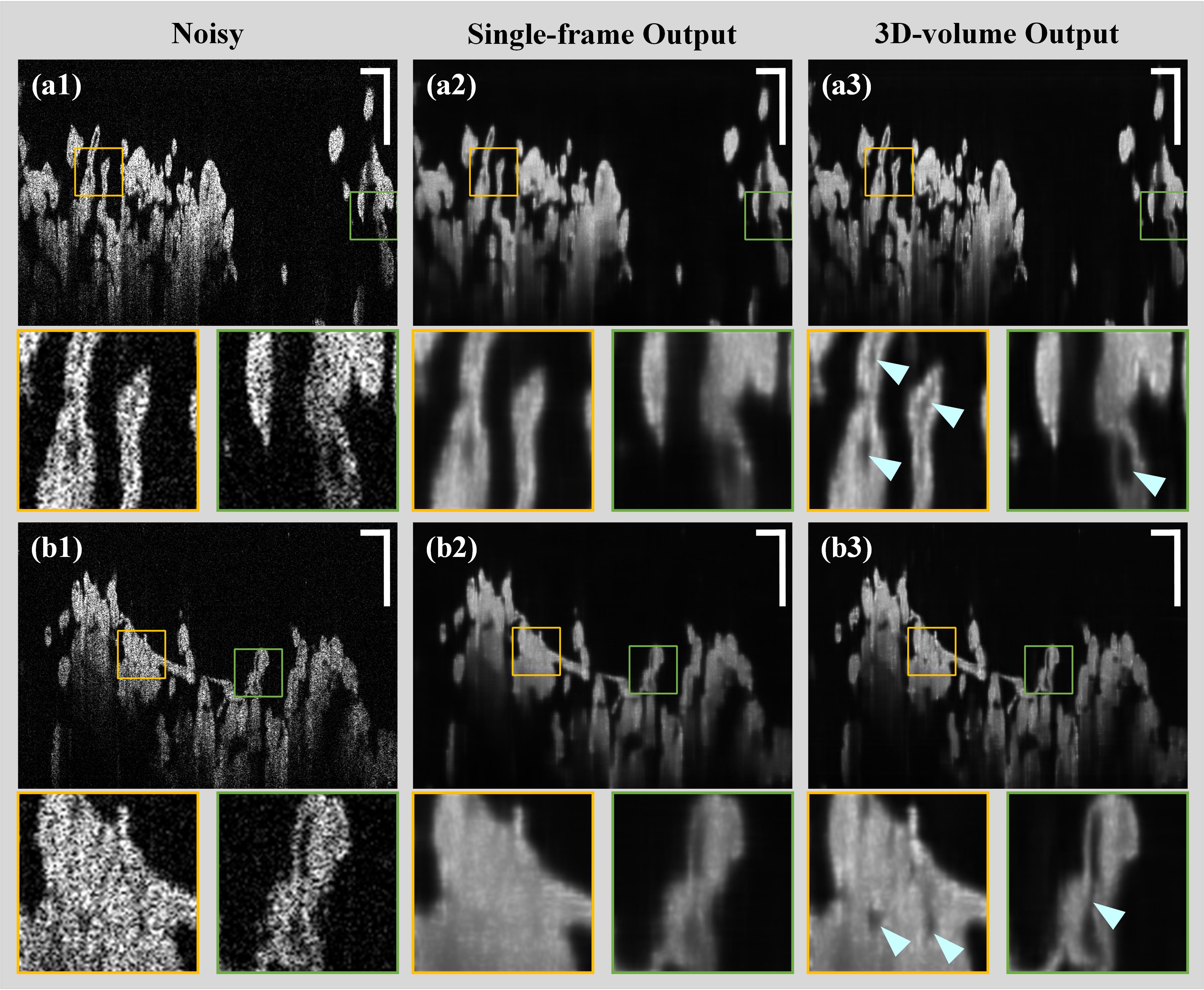}
\caption{Denoised images of the human placenta output by the proposed \emph{t}GT-OCT. Scale bar: 200 $\upmu$m.}
\label{fig:Replacenta}
\end{figure}

Fig.  \ref{fig:ReSkin} presents cropped noisy images of human skin and the corresponding denoised results of the \emph{t}GT-OCT 2D network and 3D network. Figs. \ref{fig:ReSkin}(a1) - (a3) show the skin of the inner side of the lower arm, and Figs. \ref{fig:ReSkin}(b1) - (b3) show the skin of the fingertip. As indicated by arrows, this figure reveals that the \emph{t}GT-OCT 3D network yields output images with greater image contrast and more discernible blood vessel details compared to those of the \emph{t}GT-OCT 2D network. Detailed information on the sweat duct in the epidermis and epidermal junction can be clearly observed from the image in Fig. \ref{fig:ReSkin}(b3).

Fig.  \ref{fig:Replacenta} shows the noisy images and \emph{t}GT-OCT output images of the human placenta \cite{Three‐dimensional}. Figs. \ref{fig:Replacenta}(a1) and (b1) show cropped noisy B-scans from the volumetric data, Figs. \ref{fig:Replacenta}(a2) and (b2) show denoised images output by the \emph{t}GT-OCT 2D network, and Figs. \ref{fig:Replacenta}(a3) and (c3) show denoised images output by the \emph{t}GT-OCT 3D network. After despeckling with \emph{t}GT-OCT, the shape of the placental villi and the capillaries at the end of the villi can be clearly seen in the magnified area, as indicated by arrows. Compared to the \emph{t}GT-OCT 2D network, the \emph{t}GT-OCT 3D network produces images with notably sharper vessel information, which we attribute to its superior contrast enhancement.

\begin{figure}[!h]
\centering
\includegraphics[width=0.9\linewidth]{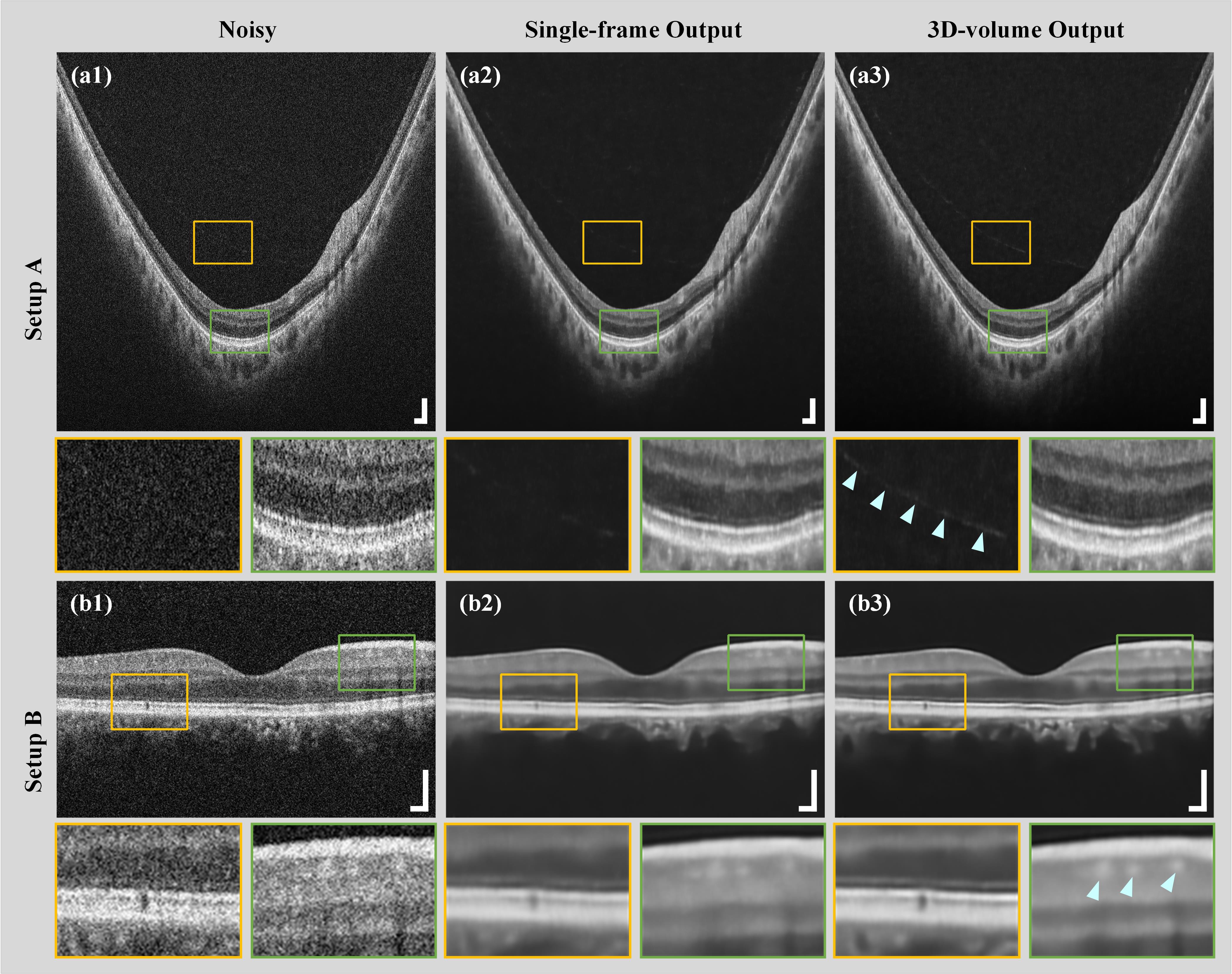}
\caption{Denoised images of the proposed \emph{t}GT-OCT obtained from the test data with different retinal OCT setups. Scale bar: 200 $\upmu$m.}
\label{fig:Resetup}
\end{figure}

\subsubsection{Generalization to different setups}
To verify the generalization capability of our approach, we tested the well-trained network on retinal images acquired from different setups. Fig. \ref{fig:Resetup}(a1) is an image from the OCT-R1 dataset obtained with a BM-400K BMizar scanner (Setup A), and Fig. \ref{fig:Resetup}(b1) is an image of the OCT-R2 dataset obtained with a Spectralis OCT system (Setup B). We ensured that the network had never seen images from the OCT-R2 dataset during training. Both the proposed \emph{t}GT-OCT 2D and 3D networks show excellent generalization capability. As shown in Fig. \ref{fig:Resetup}, speckle noise is effectively reduced, preserving fine structures and important biomarkers. The retinal layered structure remains clearly discernible after denoising, with \emph{t}GT-OCT 3D network demonstrating superior detail resolving capabilities. Consistent with the above results, the \emph{t}GT-OCT 3D network has superior detail resolution performance. The posterior vitreous detached membrane and highly reflective foci, indicated by red arrows, are notably more observable in the results of the \emph{t}GT-OCT 3D network compared to the \emph{t}GT-OCT 2D network.

\section{DISCUSSION}
To better verify the advantages of the proposed \emph{t}GT-OCT, we engaged in a comprehensive comparison with the most recently developed unsupervised OCT despeckling methodologies. Blind2Unblind (B2U) \cite{Blind2Unblind}, NBR \cite{Neighbor2Neighbor} and MAP-SNR \cite{Self-supervised} were trained on the OCT-R1 and OCT-N1 datasets to compare their denoising performance with human retina and other non-retinal samples. To ensure a fair comparison, we optimized hyperparameters across all deep learning methods to ensure t their optimal denoising performance.

\begin{figure*}[!h]
\centering
\includegraphics[width=\linewidth]{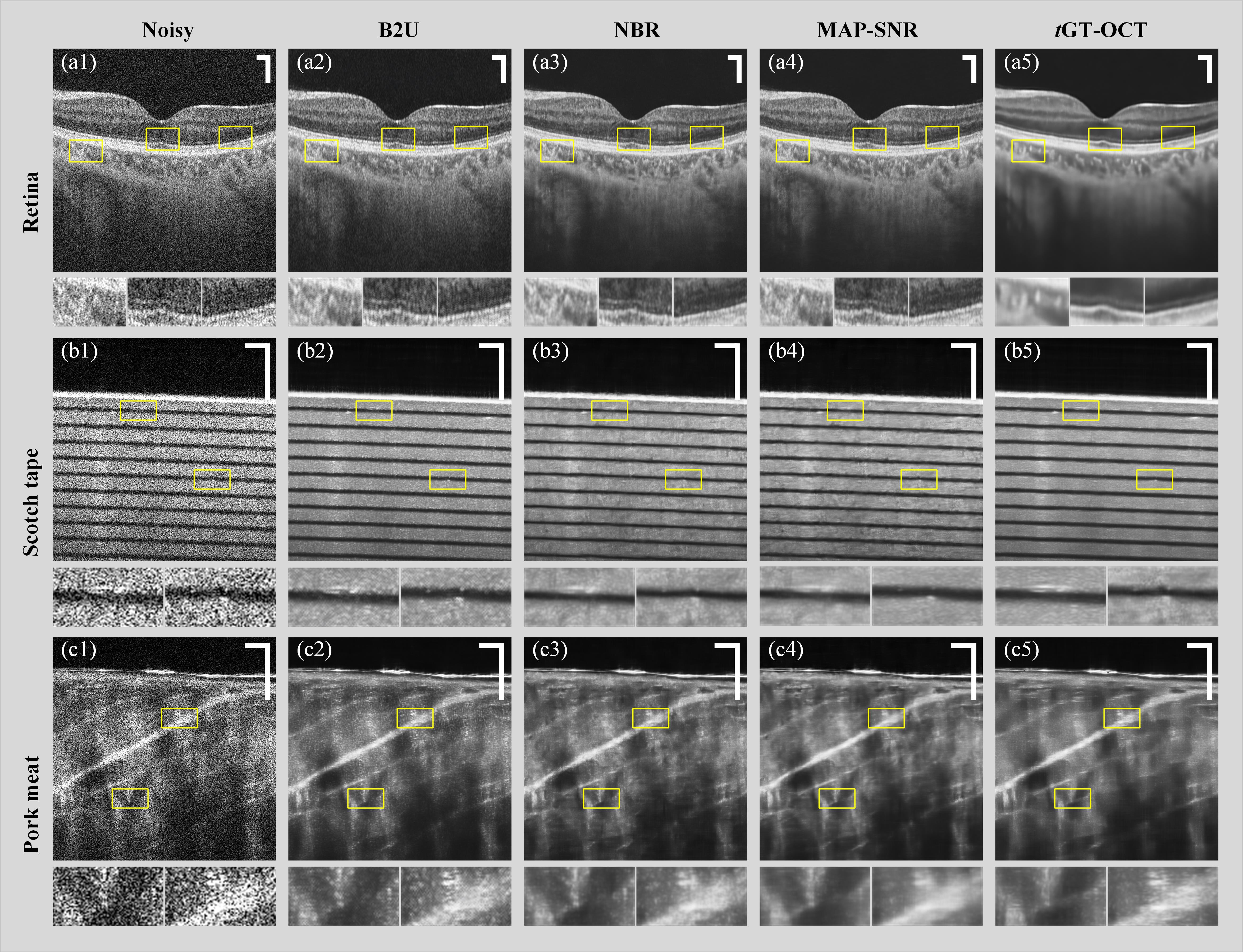}
\caption{Comparison of the denoising performance of different deep-learning-based methods with that of \emph{t}GT-OCT. Scale bar: 200 $\upmu$m.}
\label{fig:comparison}
\end{figure*}

Figs.  \ref{fig:comparison}(a1) - (a5) present the denoising results of different deep-learning-based methods and \emph{t}GT-OCT on the OCT-R1 test data. These results show that the B2U can suppress some noise while retaining some detailed information, although notable residual noise patterns are present. The NBR and MAP-SNR methods achieve a more substantial noise reduction but have compromised image quality due to the introduction of oversmoothing artifacts. In contrast, our proposed \emph{t}GT-OCT excels in despeckling performance, adeptly suppressing speckle noise while preserving detailed information and even resolving microstructures within the choroidal vascular region. 
Fig.  \ref{fig:comparison}(b1) shows the original noisy Scotch tape image, while Figs. \ref{fig:comparison}(b2) - (b5) show the denoised images obtained by various deep-learning based methods and \emph{t}GT-OCT. It is evident that B2U’s output images still contain remnants of speckle patterns, thereby resulting in poor image quality. The NBR and MAP-SNR methods suppress more noise but cause a smoothing effect that compromises the representation of certain structural details. The proposed \emph{t}GT-OCT can effectively suppress speckle noise and preserve detailed structures. 
We also compared these approaches using the image of pork, as shown in Figs. \ref{fig:comparison}(c1) - (c5). Similar to the results of Scotch tape, substantial speckle noise remains in the denoised image output by B2U and structural details are smoothed in the output image of NBR and  MAP-SNR. In the magnified regions on the right side of Figs. \ref{fig:comparison}(c2) - (c4), the pork strips are disconnected, although they are continuous in Fig. \ref{fig:comparison}(c5). It is evident that \emph{t}GT-OCT not only achieves effective denoising and microstructure resolution but also excels in generalization ability.

Table \ref{tab:metricComparison} shows no-reference evaluation metrics SNR, CNR and ENL values of the proposed \emph{t}GT-OCT and other unsupervised OCT despeckling methodologies. \emph{t}GT-OCT has the highest SNR, CNR and ENL scores in differenr sample images of dataset OCT-R1 and OCT-N1.
\begin{table*}[ht]
\centering
\caption{Quantitative comparison of the proposed \emph{t}GT-OCT and different denoising methods on different samples}
\begin{tabular}{lllllll}
\hline
\makebox[0.1\textwidth][l]{   }                         & \makebox[0.08\textwidth][l]{   }                         & \makebox[0.08\textwidth][l]{Noisy}  & \makebox[0.08\textwidth][l]{B2U}     & \makebox[0.08\textwidth][l]{NBR}      & \makebox[0.08\textwidth][l]{MAP-SNR}  & \makebox[0.08\textwidth][l]{\emph{t}GT-OCT}  \\ \hline
\multicolumn{1}{l|}{\multirow{3}{*}{Human Retina}} & SNR & 12.183 & 30.630  & 42.391   & 45.093   & \textbf{48.214}   \\
\multicolumn{1}{l|}{}                              & CNR & 1.628  & 4.085   & 4.756    & 4.602    & \textbf{5.817}    \\
\multicolumn{1}{l|}{}                              & ENL & 1.064  & 61.063  & 1032.887 & 1964.520 & \textbf{3258.357} \\ \hline
\multicolumn{1}{l|}{\multirow{3}{*}{Scotch Tape}}  & SNR & 17.795 & 39.490  & 42.159   & 40.678   & \textbf{42.994}   \\ \multicolumn{1}{l|}{}  &
CNR & 2.463  & 6.248   & 6.673    & 6.624    & \textbf{6.718}    \\
\multicolumn{1}{l|}{}                              & ENL & 0.393  & 57.594  & 117.019  & 69.625   & \textbf{139.210}  \\ \hline
\multicolumn{1}{l|}{\multirow{3}{*}{Pork Meat}}    & SNR & 16.206 & 42.247  & 41.189   & 39.081   & \textbf{43.895}   \\
\multicolumn{1}{l|}{}                              & CNR & 1.837  & 5.694   & 6.070    & 6.261    & \textbf{6.568}    \\
\multicolumn{1}{l|}{}                              & ENL & 0.342  & 128.632 & 115.372  & 59.551   & \textbf{193.111}  \\ \hline
\end{tabular}
\label{tab:metricComparison}
\end{table*}

\section{CONCLUSION}
Here we have introduced a novel speckle-free OCT imaging strategy that employs an unsupervised 3D deep learning network to distinguish and extract speckle patterns in OCT 3D volumetric data. This approach leveraged the power of the 3D convolutional network and OCT 3D imaging features, negating the necessity for clean images during training. Furthermore, our strategy maximized efficiency by incorporating a knowledge distillation mechanism to train the 2D convolutional network to achieve comparable denoising capabilities, effectively minimizing model complexity and computational demands. Experimental results with different sample images demonstrated that the proposed \emph{t}GT-OCT can clarify and reveal structures that are otherwise obscured or undetectable while preserving spatial resolution; this was achieved by fully using the global information inherent within OCT 3D volumetric data. The presented comparative and generalization studies showed that the proposed \emph{t}GT-OCT can effectively reduce speckle noise in OCT images of different samples and outperforms other deep learning methods, even achieving similar performance SM-OCT. A distilled 2D network, boasting commendable denoising performance and compact file size, stands poised for deployment in actual clinical applications. Meanwhile, this work also leverages a new perspective for studying OCT speckle-free imaging with OCT 3D imaging features and unsupervised 3D deep-learning processing.

\begin{backmatter}
\bmsection{Funding} National Natural Science Foundation of China (61905036); China Postdoctoral Science Foundation (2019M663465, 2021T140090); Fundamental Research Funds for the Central Universities (University of Electronic Science and Technology of China) (ZYGX2021J012); Medico-Engineering Cooperation Funds from University of Electronic Science and Technology of China (ZYGX2021YGCX019).


\end{backmatter}










\end{document}